# On corrected formula for graphene quantum conductivity

## N.E. Firsova[1,2]

[1]Institute for Problems in Mechanical Engineering RAS,
Bolshoy Prospect V.O. 61, St. Petersburg 199178, Russia,

[2]Peter the Great Saint Petersburg Polytechnic University,
Politechnicheskaya 29, St. Petersburg 195231, Russia

**Abstract.** - Graphene membrane irradiated by weak activating periodic electric field in terahertz range is considered. The corrected formula for the graphene quantum conductivity is found. The obtained formula gives complex conjugate results when radiation polarization direction is clockwise or it is opposite clockwise. The found formula allows us to see that the graphene membrane is an oscillating contour. Its eigen frequency coincides with a singularity point of the conductivity and depends on the electrons concentration. So the graphene membrane could be used as an antenna or a transistor and its eigen frequency could be tuned by doping in a large terahertz-infrared frequency range. The obtained formula allows us also to calculate the graphene membrane quantum inductivity and capacitance. The found dependence on electrons concentration is consistent with experiments. The method of the proof is based on study of the time-dependent density matrix. The exact solution of von Neumann equation for density matrix is found for our case in linear approximation on the external field. On this basis the induced current is studied and then the formula for quantum conductivity as a function of external field frequency and temperature is obtained. The method of the proof suggested in this paper could be used to study other problems. The found formula for quantum conductivity can be used to correct the SPPs Dispersion Relation and for the description of radiation process. It would be useful to take the obtained results into account when constructing devices containing graphene membrane nanoantenna. Such project could make it possible to create wireless communications among nanosystems. This would be promising research area of energy harvesting applications.

**1.Introduction.** - After the graphene discovery [1-2] there appeared many interesting results which were studied in the framework of 2D (flatland) model (see for instance review [3]). However it was shown 70 years ago by Peierls [4-5] and Landau [6] that strictly speaking 2D crystal cannot exist since thermal fluctuations should destroy its long range order by thermal fluctuations. After a while it was found that actually the monolayer graphene surface was always strongly corrugated [7]. In [8] it was shown that the in-plane and out-of-plane deformations mutual interaction save membrane from being crumpled.

This surface curvature under external time depending field generates "breathing" metrics which results in certain new effects. These effects proved to be essential and were for the first time analyzed in [9-11]. It was shown in [9] that the quality factor of the graphene nanoresonator considerably depends on graphene membrane metrics generating new loss mechanism based on the so called synthetic electric fields. These fields arise due to the "breathing" graphene membrane metrics activated by the external periodic electric field. In [10-11] graphene electromagnetic response was studied in terahertz range. It was shown that the induced currents were curved, the curvature essentially depending on the graphene membrane surface metrics. Results of [10-11] explain experimental data. The methods used there were based on the modified Boltzmann equation and on taking into consideration synthetic electric fields. Nowadays investigation of feature-rich electronic properties in rippled graphene becomes more and more important, [12], especially in connection with the appearance of new 2D-materials (silicone, germanene, stanene, phosphorene and so on).

Our study in [10-11] was done in the framework of quasiclassics. Now our goal is to examine graphene surface corrugations influence in nonstationary processes using quantum mechanics approach which seems to be more adequate in the case of graphene membrane. In particular in [13] nonstationary density matrix method was used. Such approach to the nonstationary physics in graphene when one takes into account corrugations influence is expected to allow us to find out new quantum effects and to obtain more adequate picture of graphene than the strain model based on elastic theory does (see [14] for details).

In [13] our problem was studied in the framework of the flatland model as an unperturbed problem leaving the perturbed case (which would require taking into account the corrugations contribution) for future research. As a first step in [13] also e-e-interactions and any kind of loss were neglected. It proved to be that the consideration even of this simplified unperturbed problem using density matrix method gave essential corrections to the physical picture compared to earlier papers (see [15-16]). So in [15] Eq.(5.13) the following formula for quantum conductivity was obtained

$$\sigma_{GSC}(\omega) = 2i\frac{e^2}{\hbar}\left[\frac{|\mu|}{\hbar\omega} + \frac{1}{4}\ln\frac{2|\mu|-\hbar\omega}{2|\mu|+\hbar\omega}\right] \quad (1)$$

in the framework of the flatland model for the case neglecting loss and e-e interactions. Here $\omega$ is the external field frequency and $\mu$ is chemical potential. In [13] for the similar case the following formula was announced

$$\sigma_{FF}(\omega) = -i\pi\frac{e^2}{\hbar}\left[\frac{E_F}{\hbar\omega} + \frac{1}{4}\ln\left|\frac{2E_F-\hbar\omega}{2E_F+\hbar\omega}\right|\right] \quad (2)$$

where $E_F$ is Fermi energy. We shall compare the results (1) and (2). The opposite sign in (2) is essential for physical reasons. Let us transform (2) to

$$\sigma_{FF}(\omega) = -2i\sigma_0\left[\frac{k_F v_F}{\omega} + \frac{1}{4}\ln\left|\frac{1-\frac{\omega}{2k_F v_F}}{1+\frac{\omega}{2k_F v_F}}\right|\right]$$

where $k_F$ and $v_F$ are Fermi wave number and Fermi velocity. For terahertz range $\omega/(2k_F v_F) \ll 1$ and consequently we have

$$\sigma_{FF}(\omega) \approx (i\omega L_{FF})^{-1} + i\omega C_{FF} \quad (3)$$

where

$$L_{FF} = (2k_F v_F \sigma_0)^{-1} = \hbar/(\pi e^2 k_F v_F) \quad (4)$$

$$C_{FF} = \frac{\sigma_0}{2k_F v_F} = \frac{\pi e^2}{4\hbar k_F v_F} = \frac{\pi e^2}{4E_F} \tag{5}$$

Note that from the obtained formula (5) we can find the quantum capacitance of the mean graphene membrane area occupied by one electron

$$c_{FF} = \sqrt{\pi n}\, e^2/(4\hbar v_F) \tag{6}$$

Note also that if condition $\omega/(2k_F v_F) \ll 1$ is not fulfilled (which we have for low electrons concentrations) then the quantum capacitance would depend on frequency as follows (see (2),(3))

$$C_{FF} = C_{FF}(\omega) = \frac{\sigma_0}{2}\ln\left|1 + 2\frac{\hbar\omega}{2E_F - \hbar\omega}\right| > 0 \tag{7}$$

while $L_{FF}$ does not depend on $\omega$. From (3) we see that graphene membrane is a nanoantenna where quantum inductivity $L_{FF}$ and capacitance $C_{FF}$ are joined in a parallel way. Remark that in [17] the graphene membrane quantum capacitance was first measured. The formula for quantum capacitance based on a two-dimensional free electron gas model was also suggested in [17]. The expression obtained in [17] coincides with (5) multiplied by $8/\pi$. Numerical calculations based on the random phase approximation of the quantum capacitance and the spin and circular -symmetric Landau parameter of the two-dimensioned fluid of massless Dirac fermions in a doped graphene sheet were presented in [18]. The measured capacitance dependence on electrons is consistent with theoretical one.

Note also that in [19] the suggestion was made that electrons in graphene must exhibit a nonzero mass when collectively excited. Using this notion the inertial acceleration of the electron collective mass and phase delay of the resulting current were considered. On the basis of this model the so-called kinetic inductance representing the reluctance of the collective mass to accelerate was introduced, calculated and measured. The obtained expression coincides with the formula (4) multiplied by $\pi^2$. Analyzing the formula for inductivity we see that $L_{FF} \sim n^{-1/2}$ that means that the less is the electrons concentration the more is the inductivity. This dependance was observed in experiment [19].

From (2) we see that graphene membrane eigen frequency (conductivity singularity) is as follows

$$\omega_{FF} = 2E_F/\hbar = 2k_F v_F = 2v_F\sqrt{\pi n}, \qquad \lim_{\omega \to \omega_{FF}} \sigma(\omega) = \infty \tag{8}$$

We see also from the formulae (5), 6) that for the found resonant frequency we have the usual equation

$$\omega_{FF} = 1/\sqrt{L_{FF} C_{FF}} \tag{9}$$

which means that graphene membrane is an oscillation contour with eigen frequency $\omega_{FF}$. Remark that the conductivity singularity in the point $\omega_{FF}$ is logarithmic instead of the pole-type we used to see in 3D. We think that it is 2D-effect.

Note that the formulae for quantum inductivity and capacitance from [17-19] actually were obtained on the basis of different models. So if we used these formulae from [17-19] to find eigen frequency according to (9) we would get the value where the conductivity (2) does not have singularity. It means that the results obtained in [17-19] do not give possibility to find correctly graphene membrane eigen frequency.

Analyzing the same way the formula (1) we see that it describes the oscillating contour with $L_{GSC} < 0$ and $C_{GSC} < 0$. So the formula (1) gives another physical picture.

In the present paper we obtain the corrected formula for graphene quantum dynamic conductivity. For external activating electric field $\vec{E}_0 e^{i\omega t}$ i.e. for the case of anticlockwise direction of polarization we get the corrected formula for conductivity for low temperatures

$$\sigma_{NF}^+ = \frac{e^2}{\hbar}\left\{\frac{\pi^2}{4}\chi(\omega) - i\pi\left[\frac{E_F}{\hbar\omega} + \frac{1}{4}\ln\left|\frac{\hbar\omega - 2E_F}{\hbar\omega + 2E_F}\right|\right]\right\} = \frac{\pi^2}{4}\frac{e^2}{\hbar}\chi(\omega) + \sigma_{FF}, \tag{10}$$

$$\chi(\omega) = \begin{matrix}1, & \omega \in [0, 2E_F/\hbar] \\ 0, & \omega > 2E_F/\hbar\end{matrix}.$$

For the case of opposite (clockwise) direction of the polarization i.e. for the external activating field $\vec{E}_0 e^{-i\omega t}$ we get

$$\sigma_{NF}^- = \frac{e^2}{\hbar}\left\{\frac{\pi^2}{4}\chi(\omega) + i\pi\left[\frac{E_F}{\hbar\omega} + \frac{1}{4}\ln\left|\frac{\hbar\omega - 2E_F}{\hbar\omega + 2E_F}\right|\right]\right\} = \frac{\pi^2}{4}\frac{e^2}{\hbar}\chi(\omega) + \frac{\pi}{2}\sigma_{GSC}. \tag{11}$$

As far as the formulae for inductivity and capacitance for clockwise polarization $\vec{E}_0 e^{-i\omega t}$ have opposite sign we get for this case (and for (1) as well) the same inductivity $L_{FF}$ and capacitance $C_{FF}$ as for the field $\vec{E}_0 e^{i\omega t}$ (see (4), (5)). Hence we see that graphene membrane can be used as a nanoantenna radiating a frequency which can be tuned in certain range. From (10), (11) we see also that changing the direction of the polarization of the activating electric field to the opposite one we get the conductivity complex conjugate to the previous one. From (10), (11) we get once again the known result for the minimal graphene conductivity $\sigma_{min} = \frac{\pi^2}{4}\frac{e^2}{\hbar}$ (see [20]).

The method of our proof will be based on the consideration of the density matrix equation in linear approximation. We shall find its exact solution for zero temperature $T = 0$ and use it for the proof of the main result. The developed construction could be used to study other problems.

**2. Statement of the problem.** - We consider the current induced in graphene membrane irradiated by the weak external complex activating electric field $\vec{E}_0 e^{i\omega t}$. We assume this field to be homogeneous which is reasonable as we consider terahertz frequency range (in terahertz frequency range the wavelength is usually much larger than graphene membrane size). We solve the problem using the quantum approach and density matrix. The Hamiltonian has a form

$$\hat{H} = \hat{H}_0 + \hat{H}_{int}, \tag{12}$$

where

$$\hat{H}_0(\vec{k}) = \hbar v_F \vec{k} \cdot \vec{\sigma} = \hbar v_F (k_1 \hat{\sigma}_1 + k_2 \hat{\sigma}_2) = \hbar v_F k \begin{pmatrix} 0 & e^{-i\varphi} \\ e^{i\varphi} & 0 \end{pmatrix}, \tag{13}$$

$$\hat{\sigma}_1 = \begin{pmatrix} 0 & 1 \\ 1 & 0 \end{pmatrix}, \quad \hat{\sigma}_2 = \begin{pmatrix} 0 & -i \\ i & 0 \end{pmatrix}, \quad \vec{k} = (k_1, k_2), \quad k_1 = k\cos\varphi, \quad k_2 = k\sin\varphi \tag{14}$$

and

$$\hat{H}_{int}(\vec{k}) = -(ev_F/c)(A_1\hat{\sigma}_1 + A_2\hat{\sigma}_2) \tag{15}$$

Here $\vec{A} = (A_1, A_2)$ is a vector potential related to the external electric field $\vec{E}_0 e^{i\omega t}$ that is

$$-c^{-1}\partial\vec{A}/\partial t = \vec{E}_0 e^{i\omega t}, \quad \vec{A} = -(c\, e^{i\omega t}/(i\omega))\vec{E}_0, \quad A_{1,2} = -(c\, e^{i\omega t}/(i\omega))\, E_{01,02}. \tag{16}$$

$$E_{01} = E_0 \cos\theta, \qquad E_{02} = E_0 \sin\theta$$

Substituting vector potential coordinates (16) into (15) we get

$$\hat{H}_{int}(\vec{k}) = \frac{ev_F}{i\omega} e^{i\omega t}[E_{01}\hat{\sigma}_1 + E_{02}\hat{\sigma}_2] = \frac{ev_F}{i\omega} E_0 e^{i\omega t} \begin{pmatrix} 0 & e^{-i\theta} \\ e^{i\theta} & 0 \end{pmatrix} \tag{17}$$

We shall calculate the induced current using the density matrix $\hat{\rho}$ which is a solution of the von Neumann equation

$$d\hat{\rho}/dt = -(i/\hbar)[\hat{H}, \hat{\rho}]. \tag{18}$$

We consider the density matrix equation (18) in linear approximation for the case of the activating field $\vec{E}_0 e^{i\omega t}$. Such approximation is reasonable as we consider weak activating electric field. The linear approximation of the equation (18) will be as follows

$$d\hat{\rho}_{int}/dt = -(i/\hbar)([\hat{H}_0, \hat{\rho}_{int}] + [\hat{H}_{int}, \hat{\rho}_0]) \tag{19}$$

Here $\hat{\rho}_0$ is a Fermi-Dirac distribution corresponding to the unperturbed case and $\hat{\rho}_{int} = \hat{\rho} - \hat{\rho}_0$ is an unknown matrix.

The mean values of the coordinates of the induced current operator $\vec{j} = (\hat{j}_1, \hat{j}_2)$, $\hat{j}_{1,2} = ev_F\hat{\sigma}_{1,2}$, will be equal to

$$j_{1,2} = ev_F Sp(\hat{\rho}\hat{\sigma}_{1,2}) \tag{20}$$

And the total current coordinates will be

$$j_{1,2}^{tot} = \int_0^\infty k\,dk \int_0^{2\pi} j_{1,2}\,d\varphi \tag{21}$$

We shall find the exact solution for the density matrix equation in linear approximation (19). Using the obtained formula we get the induced current from (20)-(21). Hence we find the formula for quantum conductivity (neglecting e-e interactions and any kind of loss mechanisms). We think that the obtained formulae after generalization with taking into account the loss, e-e interactions and inevitably existing corrugations will be useful in applications. The taking into consideration the obtained results when constructing devices containing graphene membranes would allow to increase the precision of measurements.

**3. Exact solution of the von Neumann equation in linear approximation.** - We consider the density matrix equation in linear approximation (19) for the case of the activating field $\vec{E}_0 e^{i\omega t}$.

To find the unknown matrix $\hat{\rho}_{int} = \hat{\rho} - \hat{\rho}_0$ we shall transform Hamiltonian $\hat{H}_0$ to the diagonal form. It is not difficult to check that

$$\hat{H}_0 = \hbar v_F k \hat{U}^{-1} \hat{\sigma}_3 \hat{U} \tag{22}$$

where

$$\hat{\sigma}_3 = \begin{pmatrix} 1 & 0 \\ 0 & -1 \end{pmatrix}, \quad \hat{U} = \frac{1}{\sqrt{2}}\begin{pmatrix} e^{i\varphi} & 1 \\ 1 & -e^{-i\varphi} \end{pmatrix}, \quad \hat{U}^{-1} = \frac{1}{\sqrt{2}}\begin{pmatrix} e^{-i\varphi} & 1 \\ 1 & -e^{i\varphi} \end{pmatrix} \tag{23}$$

Then we have

$$\hat{\rho}_0 = \hat{U}^{-1} \hat{f}\, \hat{U} = \hat{U}^{-1}\begin{pmatrix} f_{FD}(\varepsilon) & 0 \\ 0 & f_{FD}(-\varepsilon) \end{pmatrix}\hat{U}, \tag{24}$$

$$f_{FD}(\pm\varepsilon) = [1 + exp\{\beta(\pm\varepsilon - \mu)\}]^{-1} = [1 + exp\{\beta(\pm\hbar v_F k - \mu)\}]^{-1}, \quad \beta = \frac{1}{k_B T}. \tag{25}$$

Now we solve the equation (19) where $\hat{\rho}_{int}$ is unknown. We seek the solution in the form

$$\hat{\rho}_{int} = e^{-(i/\hbar)\hat{H}_0 t}\hat{\mathfrak{p}}_{int}e^{(i/\hbar)\hat{H}_0 t} \tag{26}$$

Substituting (26) into the equation (19) we obtain

$$d\hat{\mathfrak{p}}_{int}/dt = -(i/\hbar)[\hat{\mathcal{H}}_{int}, \rho_0] \tag{27}$$

where

$$\hat{\mathcal{H}}_{int} = e^{(i/\hbar)\hat{H}_0 t}\hat{H}_{int}\, e^{-(i/\hbar)\hat{H}_0 t} \tag{28}$$

Integrating (27) we find

$$\hat{\mathfrak{p}}_{int} = -(i/\hbar)[\hat{\mathbb{H}}_{int}, \hat{\rho}_0] \tag{29}$$

where

$$\hat{\mathbb{H}}_{int} = \int \hat{\mathcal{H}}_{int} dt = \int e^{(i/\hbar)\hat{H}_0 t}\hat{H}_{int}\, e^{-(i/\hbar)\hat{H}_0 t}\, dt \tag{30}$$

As far as

$$e^{\pm(i/\hbar)\hat{H}_0 t} = \hat{U}^{-1}\begin{pmatrix} e^{\pm iv_F kt} & 0 \\ 0 & e^{\mp iv_F kt} \end{pmatrix}\hat{U} \tag{31}$$

we find from (17) and (30)

$$\hat{\mathbb{H}}_{int} = \frac{ev_F}{i\omega}E_0 \hat{U}^{-1}\hat{I}\hat{U}, \quad \hat{I} = \int \hat{F}(t)dt \tag{32}$$

$$\hat{F}(t) = e^{i\omega t}\begin{pmatrix} e^{iv_F kt} & 0 \\ 0 & e^{-iv_F kt} \end{pmatrix}\hat{\theta}\begin{pmatrix} e^{-iv_F kt} & 0 \\ 0 & e^{iv_F kt} \end{pmatrix} \tag{33}$$

$$\hat{\theta} = \hat{U}\begin{pmatrix} 0 & e^{-i\theta} \\ e^{i\theta} & 0 \end{pmatrix}\hat{U}^{-1} \tag{34}$$

It is not difficult to see that, (23),

$$\hat{\theta} = \begin{pmatrix} \cos(\theta - \varphi) & ie^{i\varphi}\sin(\theta - \varphi) \\ -ie^{-i\varphi}\sin(\theta - \varphi) & -\cos(\theta - \varphi) \end{pmatrix}$$

Then we have

$$\hat{F}(t) = e^{i\omega t}\begin{pmatrix} \cos(\theta - \varphi) & i\sin(\theta - \varphi)e^{i\varphi}e^{2iv_F kt} \\ -i\sin(\theta - \varphi)e^{-i\varphi}e^{-2iv_F kt} & -\cos(\theta - \varphi) \end{pmatrix}$$

Hence and from (32) it follows

$$\hat{I} = \begin{pmatrix} \dfrac{e^{i\omega t}}{i\omega}\cos(\theta - \varphi) & i\sin(\theta - \varphi)e^{i\varphi}\alpha_+(t) \\ -i\sin(\theta - \varphi)e^{-i\varphi}\alpha_-(t) & -\dfrac{e^{i\omega t}}{i\omega}\cos(\theta - \varphi) \end{pmatrix} \quad (35)$$

$$\alpha_\pm(t) = \int e^{it(\omega \pm 2v_F k)}\,dt = \frac{e^{it(\omega \pm 2v_F k)}}{i(\omega \pm 2v_F k)}. \quad (36)$$

Now from (26), (29), (32) we obtain

$$\hat{\rho}_{int} = i\frac{ev_F E_0}{\hbar\omega} \times$$

$$\times \hat{U}^{-1}\begin{pmatrix} e^{-iv_F kt} & 0 \\ 0 & e^{iv_F kt} \end{pmatrix}\left[\hat{I}, \begin{pmatrix} f_{FD}(\varepsilon) & 0 \\ 0 & f_{FD}(-\varepsilon) \end{pmatrix}\right]\begin{pmatrix} e^{iv_F kt} & 0 \\ 0 & e^{-iv_F kt} \end{pmatrix}\hat{U}$$

Using (35) we obtain

$$\left[\hat{I}, \begin{pmatrix} f_{FD}(\varepsilon) & 0 \\ 0 & f_{FD}(-\varepsilon) \end{pmatrix}\right] =$$

$$= -i\sin(\theta - \varphi)\Delta(\varepsilon)\begin{pmatrix} 0 & e^{i\varphi}\alpha_+(t) \\ e^{-i\varphi}\alpha_-(t) & 0 \end{pmatrix}, \quad \Delta(\varepsilon) = f_{FD}(\varepsilon) - f_{FD}(-\varepsilon), \quad (37)$$

Hence we find

$$\hat{\rho}_{int} = i\frac{ev_F}{\hbar\omega}E_0 \sin(\theta - \varphi)\Delta(\varepsilon) \times$$

$$\times \hat{U}^{-1}\begin{pmatrix} e^{-iv_F kt} & 0 \\ 0 & e^{iv_F kt} \end{pmatrix}\begin{pmatrix} 0 & e^{i\varphi}\alpha_+(t) \\ e^{-i\varphi}\alpha_-(t) & 0 \end{pmatrix}\begin{pmatrix} e^{iv_F kt} & 0 \\ 0 & e^{-iv_F kt} \end{pmatrix}\hat{U}$$

And finally we obtain

$$\hat{\rho}_{int} = i\frac{ev_F}{2\hbar\omega}E_0 \sin(\theta - \varphi)\Delta(\varepsilon)\begin{pmatrix} (\tilde{\alpha}_+ + \tilde{\alpha}_-) & e^{-i\varphi}(\tilde{\alpha}_- - \tilde{\alpha}_+) \\ e^{i\varphi}(\tilde{\alpha}_+ - \tilde{\alpha}_-) & -(\tilde{\alpha}_+ + \tilde{\alpha}_-) \end{pmatrix} \quad (38)$$

$$\tilde{\alpha}_\pm(t) = e^{\mp 2iv_F kt}\int e^{i\omega t}e^{\pm 2iv_F kt}\,dt = \frac{e^{i\omega t}}{i(\omega \pm 2v_F k)}. \quad (39)$$

## 4. Calculation of the induced current and quantum conductivity.

Now we can find from (20) the first coordinate of the current

$$j_1 = ev_F Sp(\hat{\rho}\hat{\sigma}_1) = ev_F[Sp(\hat{\rho}_0\hat{\sigma}_1) + Sp(\hat{\rho}_{int}\hat{\sigma}_1)] = j_{10} + j_{1int}. \tag{40}$$

and total current coordinate (21)

$$j_1^{tot} = \int_0^\infty kdk \int_0^{2\pi} [j_{10}(k,\varphi) + j_{1int}(k,\varphi)] d\varphi \tag{41}$$

Note that

$$j_{10}(k,\varphi) = ev_F Sp\left(\hat{U}^{-1} \begin{pmatrix} f_{FD}(\varepsilon) & 0 \\ 0 & f_{FD}(-\varepsilon) \end{pmatrix} \hat{U}\hat{\sigma}_1\right) =$$

$$= ev_F[f_{FD}(\varepsilon) - f_{FD}(-\varepsilon)] \cos\varphi = ev_F \Delta(\varepsilon) \cos\varphi. \tag{42}$$

So we have

$$\int_0^\infty kdk \int_0^{2\pi} j_{10}(k,\varphi) d\varphi = ev_F \int_0^\infty \Delta(\varepsilon) kdk \int_0^{2\pi} \cos\varphi \, d\varphi = 0$$

Now it follows from (41) that

$$j_1^{tot} = \int_0^\infty kdk \int_0^{2\pi} j_{1int}(k,\varphi) d\varphi. \tag{43}$$

Let us calculate this total coordinate. From (40) we see

$$j_{1int} = ev_F Sp(\hat{\rho}_{int}\hat{\sigma}_1).$$

and consequently, (38),

$$j_{1int} = i\frac{e^2 v_F^2 E_0}{2\hbar\omega} \Delta(\varepsilon) \sin(\theta - \varphi) Sp\begin{pmatrix} e^{-i\varphi}(\tilde{\alpha}_- - \tilde{\alpha}_+) & (\tilde{\alpha}_+ + \tilde{\alpha}_-) \\ -(\tilde{\alpha}_+ + \tilde{\alpha}_-) & e^{i\varphi}(\tilde{\alpha}_+ - \tilde{\alpha}_-) \end{pmatrix}.$$

So we have

$$j_{1int} = -\frac{e^2 v_F^2 E_0}{\hbar\omega} \Delta(\varepsilon) \sin(\theta - \varphi) \sin\varphi \, (\tilde{\alpha}_+ - \tilde{\alpha}_-).$$

It is not difficult to calculate that, (39),

$$(\tilde{\alpha}_+ - \tilde{\alpha}_-) = -ie^{i\omega t}\left[\frac{1}{\omega + 2v_F k} - \frac{1}{\omega - 2v_F k}\right].$$

So we see

$$j_{1int} = i\frac{e^2 v_F^2}{\hbar\omega} E_0 e^{i\omega t}\left[\frac{1}{\omega + 2v_F k} - \frac{1}{\omega - 2v_F k}\right] \Delta(\varepsilon) \sin(\theta - \varphi) \sin\varphi.$$

As far as

$$\int_0^{2\pi} \sin(\theta - \varphi) \sin \varphi \, d\varphi = -\pi \cos \theta$$

we find from (43)

$$j_1^{tot} = -i\pi \frac{e^2}{\hbar} \frac{v_F^2}{\omega} E_{0x} e^{i\omega t} \int_0^{\infty} \Delta(\varepsilon) \left[\frac{1}{\omega + 2v_F k} - \frac{1}{\omega + i0 - 2v_F k}\right] k dk. \quad (44)$$

For zero temperature $T = 0$ we obtain

$$j_1^{tot} = -i\pi \frac{e^2}{\hbar} \frac{v_F^2}{\omega} E_{0x} e^{i\omega t} \int_0^{k_F} \left[\frac{1}{\omega + 2v_F k} - \frac{1}{\omega + i0 - 2v_F k}\right] k dk.$$

After calculation of the integral we have

$$j_1^{tot} = \frac{e^2}{\hbar} \left\{ \frac{\pi^2}{4} \chi(\omega) - i\pi \left[\frac{E_F}{\hbar\omega} + \frac{1}{4} \ln \left|\frac{\hbar\omega - 2E_F}{\hbar\omega + 2E_F}\right|\right]\right\} E_{0x} e^{i\omega t} = \sigma_{xx}^+ E_{0x} e^{i\omega t}. \quad (45)$$

The same way we could get

$$j_2^{tot} = \frac{e^2}{\hbar} \left\{ \frac{\pi^2}{4} \chi(\omega) - i\pi \left[\frac{E_F}{\hbar\omega} + \frac{1}{4} \ln \left|\frac{\hbar\omega - 2E_F}{\hbar\omega + 2E_F}\right|\right]\right\} E_{0y} e^{i\omega t} = \sigma_{yy}^+ E_{0y} e^{i\omega t}. \quad (46)$$

From (45) and (46) we get the formula (10) for $\sigma_{NF}^+ = \sigma_{xx}^+ = \sigma_{yy}^+$. The formula (11) can be proved in a similar way if we consider activating electric field $\vec{E}_0 e^{-i\omega t}$ with opposite (clockwise) direction of polarization.

So dynamic conductivities in irradiated graphene for clockwise and anticlockwise directions of polarization are complex conjugate.

Note also that using (44) we could find for irradiated graphene the quantum conductivity, inductivity and capacitance dependence on temperature.

**5.Discussion and conclusion.** - We considered graphene membrane electromagnetic response to a weak periodic electric field in terahertz-infrared range in the framework of flatland model. We found the corrected formula for graphene quantum conductivity and showed that for clockwise and anticlockwise directions of the radiation polarization conductivities would be complex conjugate. We found and studied graphene nanoantenna properties which proved to be quantum effect. To obtain the formulas for graphene quantum conductivity, inductivity and capacitence we used quantum approach based on the nonstationary density matrix method neglecting e-e interactions and any loss. We found the exact solution of the von Neumann equation (density matrix equation) in linear approximation. On this basis we obtained the induced current and quantum conductivity. For low temperatures we found simple formula. This corrected formula for quantum conductivity allowed to see that graphene membrane was actually oscillating contour and to find quantum inductivity, quantum capacitance and eigen frequency. The value of the found eigen frequency $\omega_{FF}$ coincides with the singularity of the conductivity. It means that graphene membrane is an antenna radiating this frequency or this is a resonant frequency of the graphene transistor. The method of the proof could be useful in other research areas.

It is interesting that the found eigen frequency $\omega_{FF}$ is equal to the single-particle threshold for the real part of the conductivity which can be seen from the formula for this threshold obtained in [21]. By the way we see the same threshold for the real part also in the formulae (10), (11). Note that for $T > 0$ we shall not have such stepwise jumping for the real part. It is also interesting that the found quantum conductivity has logarithmic type singularity for resonant frequency instead of the pole-type one we used to have in 3D. We think that it is 2D-effect.

As far as charge carriers concentration $n$ can vary from $10^9$ to $10^{13}$ we see from our formula (9) that using doping by constant electric field (see for instance [22]) we can vary eigen frequency approximately from 10 to 100 terahertz.

As far as our formula was obtained neglecting e-e-interactions we plan to find a corrected value $\omega_{FF}^{int}$ of the eigen frequency for the case when e-e-interactions are taken into account.

To find correct value of the eigen frequency we should also take into account the graphene surface corrugations influence. For instance for the graphene nanoresonator this lead to extra loss and lower quality factor as we demonstrated in [9]. In [10-11] where we studied induced current trajectories for corrugated graphene we found that they were curved and the curvature was determined by the graphene surface form. This result was consistent with experiment where current paths were obtained and they proved to be curved [23] which can be explained only by our results. We also found there that the induced currents have phase delay depending on the point $(x, y)$. So it may happen that when we develop our quantum approach taking into account surface corrugations influence we may find that the graphene membrane conductivity, inductivity, capacitance and eigen frequency will prove to depend on the point $(x, y)$. We hope to answer these questions in our next papers which would be useful for constructing the improved quantum description of graphene nanoantenna and its applications to wireless communications among nanosystems.